\documentclass[%
 reprint,
superscriptaddress,
nofootinbib,
 amsmath,amssymb,
 aps,
prd,
floatfix,
longbibliography
]{revtex4-1}

\usepackage{textcomp}

\usepackage{array}
\usepackage{dcolumn}
\usepackage{bm}
\usepackage{hyperref}
\usepackage{amsfonts,amssymb,stackengine,graphicx}
\setstackgap{S}{-1.5pt}

\usepackage[dvipsnames]{xcolor}
\usepackage[mathlines]{lineno}
\usepackage{tabularx}
\usepackage{subfigure}
\usepackage{multirow}
\usepackage{gensymb}
\usepackage[normalem]{ulem}
\usepackage[export]{adjustbox}
\usepackage{url}
\usepackage{academicons}
\usepackage{enumitem}
\usepackage{placeins}
\usepackage{braket}

\hypersetup{
    pdfnewwindow=true,    
    colorlinks=true,      
    linkcolor=blue,       
    citecolor=blue,       
    filecolor=blue,       
    urlcolor=blue         
}

\newcommand{\orcid}[1]{\href{https://orcid.org/#1}{\textcolor[HTML]{A6CE39}{\aiOrcid}}}

\newcolumntype{Y}{>{\centering\arraybackslash}X}

\begin{document}
\title{Boosted Dark Matter Directionality in Large Liquid Scintillator Detectors}

\author{Samuel S. H. Tse}
\email{samueltsh@link.cuhk.edu.hk}
\affiliation{
Department of Physics, The Chinese University of Hong Kong, Shatin, Hong Kong, China}
\affiliation{
Institute of High Energy Physics, Beijing, China}

\author{Qishan Liu}
\email{liuqs@ihep.ac.cn}
\affiliation{
Department of Physics, The Chinese University of Hong Kong, Shatin, Hong Kong, China}
\affiliation{
Institute of High Energy Physics, Beijing, China}
\affiliation{China Center of Advanced Science and Technology, Beijing 100190, China}

\author{Kenny C. Y. Ng}
\email{kcyng@phy.cuhk.edu.hk}
\affiliation{
Department of Physics, The Chinese University of Hong Kong, Shatin, Hong Kong, China}

\author{Koun Choi}
\email{koun@ibs.re.kr}
\affiliation{
Center for Underground Physics, Institute for Basic Science, Daejeon, Korea}

\author{Yu-Feng Li}
\email{liyufeng@ihep.ac.cn}
\affiliation{
Institute of High Energy Physics, Beijing, China}

\date{\textcolor{blue}{\today}}


\begin{abstract}

Light dark matter can be boosted by cosmic-ray–dark matter interactions, which opens the possibility of direct-detection searches with large neutrino detectors. We consider the signal produced by boosted dark matter undergoing quasi-elastic scattering off carbon and knocking out a neutron in a large liquid scintillator detector. A distinctive delayed gamma ray from neutron capture can be reconstructed after $\sim210$ $\mu \mathrm{s}$, which is paired up with the prompt signal during neutron propagation. We find that 100 kt$\cdot$yr of data from the neutron-coincidence channel will be able to probe new parameter space compared to free proton signals found in~\cite{Ema_2021}. We further consider the possibility of utilizing directional information from the displaced vertex associated with the capture of the final-state neutron. However, we find that directional information does not materially improve the sensitivity due to the large nucleon scattering angle.

\end{abstract}

\maketitle

\section{\label{sec:intro}Introduction}
\par Astrophysical and cosmological observations, including Galactic rotation curves, gravitational lensing, cosmic microwave background, and galaxy cluster dynamics, provide compelling evidence for the existence of dark matter (DM)~\cite{Cirelli:2024ssz, Bertone:2016nfn}. DM dominates the matter content of the Universe, yet its particle nature remains unknown. Numerous experiments such as direct detection~\cite{XENON:2025vwd,LZ:2022lsv,cui2017,PandaX-4T:2021bab,CDEX:2019hzn}, indirect detection~\cite{Fermi-LAT:2025gei,LHAASO:2022yxw, LHAASO:2024upb, IceCube:2025fcn}, and collider searches~\cite{ATLAS:2020uiq}, have not yielded a conclusive detection. In conventional direct detection searches, sensitivity to sub-GeV DM is limited by kinematics. Halo DM particles typically arrive at Earth with velocities of order $v \sim 10^{-3}$~\cite{lewin1996}. DM particles with sub-GeV mass yield nuclear recoil energies that are below the energy thresholds of most detectors~\cite{XENON:2025vwd,LZ:2022lsv,PandaX-4T:2021bab}. 

An alternative arises when DM is boosted to relativistic energies. Such boosted dark matter (BDM) can then trigger detectors with enhanced recoil energies, opening new detection channels in existing direct detection experiments~\cite{Bringmann_2019,Ema_2021, PandaX:2024syk,LZ:2025iaw,CDEX:2025mcb,Aalbers:2022dzr,Bardhan:2022bdg,Dent:2020crbdm,Ghosh:2024dqw}. Since DM can interact with nucleons and produce detectable signals, the cosmic-ray-boosted-dark matter (CRBDM) scenario, where energetic cosmic rays (CR) scatter off cold DM in the Galactic halo to relativistic energies~\cite{Bringmann_2019,Cappiello:2018hsu}, becomes unavoidable. As BDM arrives at the detector with enhanced nuclear recoil energy, neutrino detectors can be an even better probe because of their large target volume~\cite{Ema_2021, Cappiello:2019qsw, Agashe:2014yua,berger2015,choi2024new,Super-Kamiokande:2022ncz,dutta2024,Arguelles:2019xgp, DUNE:2020fgq,Cappiello:2024icecube}.

Large liquid scintillator detectors (LLSD) provide unique advantages for BDM searches involving nucleon or nuclear recoils. As BDM interacts analogously to neutrino neutral‑current scattering, LLSD with low energy thresholds and kiloton-scale target masses enable sensitivity to MeV-scale signals arising from BDM scattering on protons or nuclei~\cite{dutta2024, Cappiello:2019qsw}. When the BDM momentum reaches hundreds of MeV, its de Broglie wavelength becomes comparable to the nucleon size, making it possible to resolve and directly interact with individual nucleons inside the nucleus. In this regime, quasi-elastic scattering (QES) begins to dominate over interactions with the entire nucleus~\cite{Ankowski}, resulting in the knockout of nucleons from the nucleus. These interactions can be tracked in Liquid Scintillator (LS) detectors, providing a method to probe BDM.

Directional information can improve the signal-to-background ratio if the source is dominated by a particular direction compared to the background. Due to the DM profile~\cite{Navarro_1996,Burkert:1995yz, 1965TrAlm...5...87E, Hernquist:1990fh}, CRBDM should predominantly come from the Galactic center (GC) direction. Therefore, nucleons knocked out by CRBDM scattering are expected to be focused away from the GC direction. Traditionally, charged particles such as protons do not provide directional information in LLSD because of the isotropic and relatively long-lasting scintillation light (with decay times ranging from a few ns to $\mu$s~\cite{Beretta2025JUNO}). Previous work has proposed using the first-photon hitting time to trace the charged particle direction with machine-learning techniques~\cite{Yang:2023rbg}. An appealing alternative, which motivates this work, is to utilize the delayed signal from the capture of knocked-out neutrons~\cite{choi2024new,lin2025detectinglightdarkmatter}, whose interaction vertices can preserve partial directional information~\cite{qishanliu2026,Fischer:2015oma, Vogel:1999zy}. The neutron-tagging efficiency in LLSD can be close to 100\%~\cite{JUNO:2022lpc}, which is another major reason for choosing LLSD. By using neutron interactions to extract directional information, we investigate the potential improvement in constraints on spin-independent DM--nucleon scattering.

In this work, we consider BDM scattering off bound neutrons and the potential use of neutron directionality for signal‑background separation. We first calculate the BDM flux arriving at an underground LLSD and the rate of neutrons knocked out by BDM. We then simulate the propagation of neutrons in LS using Geant4 in Section~\ref{sec_2}. We evaluate the detector's sensitivity limits, with and without directional information, and compare them with previous work that used protons as targets~\cite{Ema_2021} in Section~\ref{sec_3}. In Section~\ref{sec_4}, we explain the mild improvement from directionality in this work and discuss alternative channels which may improve the sensitivity further.

\section{\label{sec_1} From BDM Flux to Event Rates}
In this section, we select a DM model and calculate the BDM flux arriving at Earth from different directions. We then model the Earth attenuation effect and compute the attenuated BDM flux that reaches the detector. Next, we calculate the BDM–nucleus scattering process that emits a neutron using the quasi-elastic regime, and determine the rate of neutrons knocked out in different directions inside the detector.
\subsection{\label{sec:BDM-flux}BDM Model and Flux}
We consider DM as a Dirac fermion $\chi$ with mass $m_\chi$ that interacts with Standard Model particles via a scalar mediator $\phi$ of mass $m_\phi$, following the framework outlined in~\cite{Ema_2021}. The differential cross section for an incoming CR particle $i$ with mass $m_i$ and kinetic energy (KE) $K_i$ scattering off DM is given by
\begin{align}
    \label{ema_cross}
    \frac{d\sigma_{i\rightarrow\chi}}{dK_\chi}
    &= \frac{1}{K^{\text{max}}_\chi}
       \frac{g_{\chi \phi}^2\,g_{n\phi}^2}{16\pi s}
       \frac{(-t + 4m_i^2)(-t + 4m_\chi^2)}{(m_\phi^2 - t)^2}\,n_i^2
    \notag \\
    &\quad \times F_i^2(-t)\,\Theta(K^{\text{max}}_\chi - K_\chi),
\end{align}
where $g_{\chi\phi}$ and $g_{n\phi}$ are coupling constants, $K_\chi$ is the KE of the outgoing boosted DM particle, $K^{\text{max}}_\chi$ is the maximum KE that DM can acquire from the interaction, $s$ and $t$ are the Mandelstam variables, $n_i$ is the number of nucleons in the CR species, $F_i(-t)$ is the nuclear form factor. In this work, we assume isospin symmetry, i.e. the mediator couples equally to protons and neutrons, $g_{n\phi}=g_{p\phi}$. For details of the theory, we refer the reader to Ref.~\cite{Ema_2021}. The coupling constants here can be related to the spin-independent DM–nucleon scattering cross section $\bar{\sigma}_n\equiv g_{\chi\phi}^2 g_{n\phi}^2 \mu_{\chi n}^2/(\pi m_{\phi}^4)$ with the reduced mass~$\mu_{\chi n}$. Given the relativistic speeds of CRs, cold halo DM can be treated as being at rest. The explicit expression for $K^{\text{max}}_\chi$ is then~\cite{Bringmann_2019}
\begin{align}
    K^{\text{max}}_\chi = \frac{K_i^2 + 2m_iK_i}{K_i + \frac{(m_i+m_\chi)^2}{2m_\chi}}.
\end{align}
We consider only two dominant CR species, $i=\{p^+, He^{2+}\}$, and adopt the simple dipole nuclear form factor used in~\cite{Bringmann_2019,Ema_2021}
\begin{align}
    \label{simple_dipole}
    F_i(Q^2) = \frac{1}{\left(1+\frac{Q^2}{\Lambda_i^2}\right)^2},
\end{align}
where $\Lambda_{p^+} = 770$ MeV and $\Lambda_{He^{2+}} = 410$ MeV~\cite{2004ADNDT..87..185A}.

To model the BDM flux, we use a cylindrical leaky‑box model of cosmic rays with a height of 2 kpc and a radius of 10 kpc~\cite{Ema_2021}. The BDM flux is given by
\begin{align}
    \label{DM_flux}
    \frac{d\Phi_\chi}{dK_\chi d\Omega_\chi} = \frac{D(b,l)}{m_\chi} \sum_i \int_{K_{i}^{\min}}^\infty  dK_i \frac{d \Phi_i}{dK_i d\Omega_\chi} \frac{d\sigma_{i\rightarrow\chi}}{dK_\chi},
\end{align}
where $\Omega_\chi$ corresponds to the BDM arrival direction relative to the detector, ${d \Phi_i}/{dK_i d\Omega_\chi} $ is the differential local interstellar (LIS) CR flux, $K_{i}^{\min}$ is the minimum KE of the CR particle required to boost the BDM to $K_\chi$
\begin{align}
    K_{i}^{\min}=\left(\frac{K_\chi}{2}-m_i \right) \left[\, 1\pm \sqrt{1+\frac{2K_\chi}{m_\chi}\frac{(m_i+m_\chi)^2}{(2m_i-K_\chi)^2}}  \, \right],
\end{align}
where the $+$ ($-$) sign is for $K_\chi > 2m_i$ ($K_\chi < 2m_i$). We assume the CR flux is isotropic and homogeneous and use interpolated data from~\cite{Zhang2015} for the low energy regime and an analytical formula adapted from~\cite{dellatorre2016observationsnearearthlocal, Boschini_2017} that applies only in the high‑energy regime (rigidity $\geq 1$ GV). However, the two curves do not align smoothly. To ensure a smoother transition between the two curves, we apply energy cutoffs to switch the models: 1.5 GeV for protons and 0.9 GeV for helium. Additionally, we set the LIS CR flux to zero when $K_i$ is less than 22 MeV for protons and 10 MeV for helium, following~\cite{Ema_2021}.
The D-factor $D(b,l)$ corresponds to the line-of-sight (l.o.s.) integral of the DM column density
\begin{align}
    \label{D_factor}
    D(b,l) = \int_{l.o.s.} dL~\rho_{\chi},
\end{align}
where $\rho_{\chi} $ is the DM density. The l.o.s. integration bounds depend on the direction specified by Galactic latitude $b$ and longitude $l$. The DM density is modeled using the Navarro–Frenk–White (NFW) profile~\cite{Navarro_1996}:
\begin{align}
    \rho_{\chi}(r) = \rho_0 \frac{r_0(r_0+r_c)^2}{r(r+r_c)^2},
\end{align}
where $r$ is the galactocentric distance, $r_0 = 8.5$ kpc is the distance from the solar system to the GC, $\rho_0 = 0.42$ GeV cm$^{-3}$ is the local DM density, and $r_c = 20$ kpc is the scale radius for Milky Way~\cite{Buch_2019,BENITO2021100826,Pato_2015}.

\subsection{\label{sec:earth-attenuation}Earth Attenuation}
LLSDs are built underground to reduce background noise from CRs. Before reaching the LLSD, BDM may scatter with the particles in the atmosphere and the Earth, losing energy and potentially changing direction. This attenuation effect can influence the BDM flux by shifting the energy spectrum to lower energies. Compared to the Earth, the attenuation effect from the atmosphere is negligible. Depending on the properties of BDM, this shift can both facilitate detection (by bringing the BDM energies into the targeted range) and pose challenges (by excessively slowing down the BDM to energies that cannot trigger any observable signal)~\cite{Emken_2018}. 

Following~\cite{Bringmann_2019, Ema_2021}, we approximate the energy loss at each scattering by its average value and assume there is no change in the BDM direction after each scattering. The average attenuated BDM KE, $\bar{K}_\chi$, after traveling a distance $z$ through the Earth, satisfies the following differential equation:
\begin{align}
\label{attenuated_ODE}
\frac{d\bar{K}_\chi}{dz} = -n_B \int dK_B \, K_B \frac{d\sigma_{\chi \rightarrow B}}{dK_B},
\end{align}
where $n_B$ is the number density of the Earth’s constituent particles. The differential cross section $\frac{d\sigma_{\chi \rightarrow B}}{dK_B}$ is modified by exchanging the indices in Eq.~(\ref{ema_cross}), as it now describes energetic BDM scattering off Earth particles. We assume that the Earth consists solely of protons and neutrons, neglecting nuclear binding effects~\cite{Ema_2021}. Following~\cite{Ema_2021,Emken_2018}, we ignore nuclear form factors for protons and neutrons in the attenuation calculation as a conservative approximation. By solving Eq.~(\ref{attenuated_ODE}), we obtain $\bar{K}_\chi(K_\chi, z)$ with the initial condition $\bar{K}_\chi(K_\chi, 0) = K_\chi$. 

\begin{figure}
    \centering
    \includegraphics[width=\linewidth]{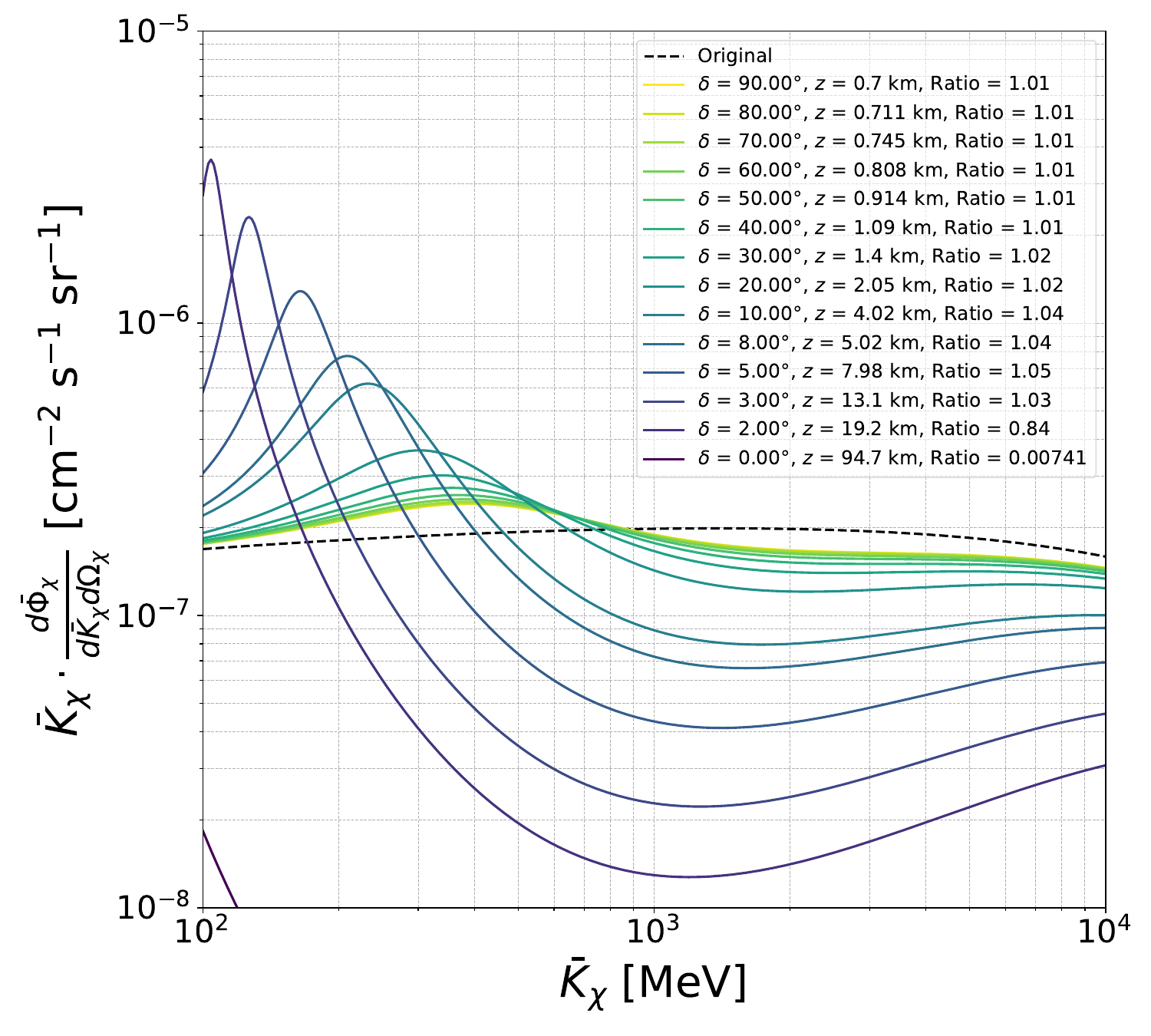}
    \caption{The BDM flux spectrum at different $\delta$ originating from $b = 0.1\degree$ and $l = 0.1\degree$, for both the unattenuated (black) and attenuated (colored) cases, with $\bar{\sigma}_n = 3.71 \times 10^{-34}~\mathrm{cm}^2$, $m_\chi = 1~\mathrm{MeV}$, and $m_\phi = 1~\mathrm{GeV}$. 
    The ratio shown in the legend is the attenuated-to-unattenuated flux ratio, integrated over $\bar{K}_\chi = 100~\mathrm{MeV}$ to $10~\mathrm{GeV}$.}
    \label{fig:attenuated_flux}
\end{figure}

For detector altitudes $\delta \geq 0^{\circ}$, we take the Earth's mass density as $2.7$ g/cm$^{3}$ since the detector is closer to the surface~\cite{Ema_2021, Emken_2018, BetancourtKamenetskaia:2025noa}. For $\delta < 0 ^{\circ}$, we take the Earth's average mass density to be $5.5$ g/cm$^{3}$. Note that this attenuation model accounts only for the energy shift in the spectrum, as the total flux is conserved. Hence, the relationship between the attenuated BDM flux and the initial BDM flux is given by: 
\begin{align}
\label{flux_conserve}
\frac{d\bar{\Phi}_\chi}{d\bar{K}_\chi d\Omega_\chi} \, d\bar{K}_\chi =\frac{d\Phi_\chi}{dK_\chi d\Omega_\chi} d K_\chi.
\end{align}
As an example, the unattenuated and attenuated BDM fluxes at $(b=0.1\degree, l=0.1\degree)$ for a fixed set of parameters at different altitudes for an LLSD located 700 m underground are illustrated in Fig.~\ref{fig:attenuated_flux}. While $z$ increases as $\delta$ decreases, it grows rapidly once $\delta \leq 0$. This attenuation model is treated as a conservative method, as it overestimates the Earth's stopping power compared to simulation results~\cite{Emken_2018, Cappiello:2019qsw}. 

\begin{figure*}[!t]
    \centering
    \includegraphics[width=\textwidth]{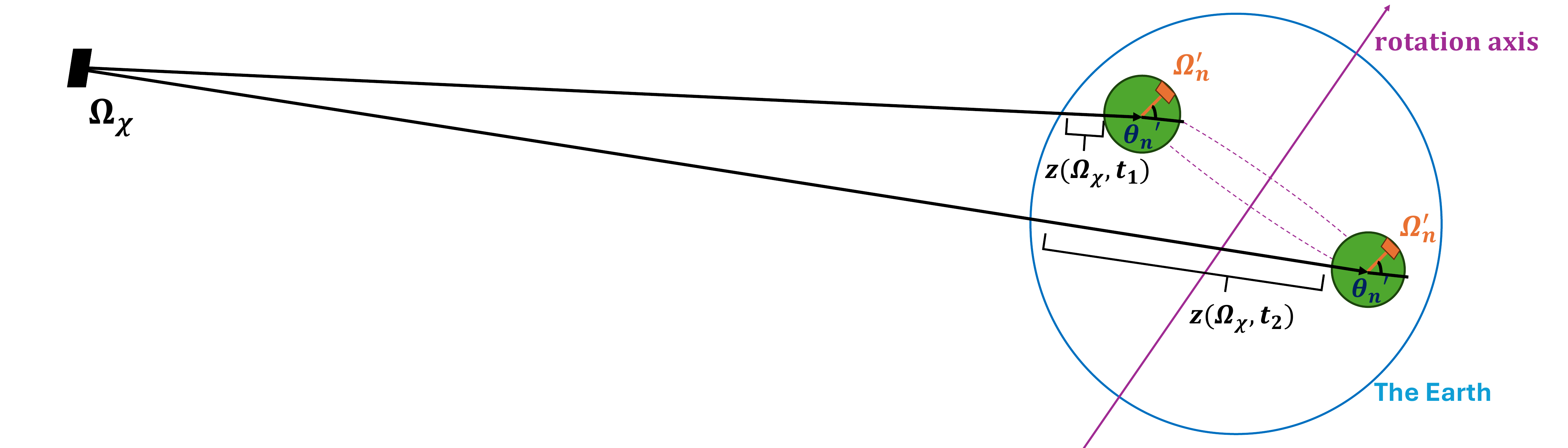}
    \caption{The detector's orbit due to Earth's rotation (dotted ring in purple) affects the depth $z$ that BDM must travel before reaching the detector from a given sky patch $\Omega_\chi$ at two different times, $t_1$ and $t_2$. The LLSD (green circles) is taken to be located in the Northern Hemisphere at a latitude of approximately $22^\circ$\,N and about $700~\mathrm{m}$ underground. The depth $z$ depends on both the direction of the sky patch $\Omega_\chi$ and the time $t$. This depth influences the attenuated BDM KE $\bar{K}_\chi$, which in turn affects the cross section. For a given directional patch in the detector frame $\Omega_n'$, both $\Omega_n'$ and $\theta_n'$ are time-independent because the Earth's diameter is negligible compared to the distance of the source.}
    \label{fig:sky_patch}
\end{figure*}

\subsection{\label{Neutrons from BDM-Nucleus Scattering} Neutrons from BDM-Nucleus Scattering}
This work targets interactions in which a BDM particle knocks out a neutron, allowing us to probe directionality. The most abundant element that contains neutrons in organic LS is ${}^{12}\mathrm{C}$. The BDM particles $\chi$ arrive at the detector and can knock out neutrons $n$ via the process
\begin{align}
    \label{reaction_EQ}
    \chi + {}^{12}\mathrm{C} \rightarrow \chi + n + {}^{11}\mathrm{C}^*,
\end{align}
where ${}^{11}\mathrm{C}^*$ is the excited residual nucleus.

For sufficiently large momentum transfer, QES with impulse approximation~(IA) provides an excellent description of the cross section for the reaction shown in Eq.~(\ref{reaction_EQ}). Following~\cite{Liangliang2023,lin2025detectinglightdarkmatter}, we make the change of variables $(E_\chi',\Omega_\chi') \rightarrow (K_n',\Omega_n')$ through a Jacobian transformation and obtain the DM–carbon QES differential cross section
\begin{align}
    \frac{d^2\sigma_{\chi \rightarrow C}}{dK_n' \, d\Omega_n'} 
    &= (A-Z) \frac{g_{\chi \phi}^2 g_{n\phi}^2}{16\pi^2} 
       \int d^3\vec{p}\, dE_R \,P_{\mathrm{RFG}}(\vec{p},E_R) \,  \notag \\
    &\quad \times \frac{1}{4 E_n E_\chi'} \frac{|\vec{p'}|}{|\vec{k}|}\frac{(4 m_\chi^2 + Q^2)(4 m_n^2 + \tilde{Q}^2)}{(m_\phi^2 + Q^2)^2} 
       \, F_n^2(Q^2) \notag \\
    &\quad \times  \,
       K_{\mathrm{Pauli}}(|\vec{q}|) \,\delta\!\left(E_\chi -E_\chi' + E_n - E_n' - E_R\right) ,
\end{align}
where $A=12$ and $Z=6$ are the mass number and atomic number of the carbon nucleus, $\Omega_n'$ and $K_n'$ are the direction and KE of the knocked-out neutron, $k = (E_\chi, \vec{k})$ and $k' = (E_\chi', \vec{k}')$ are the four-momenta of the BDM before and after scattering in the carbon nucleus rest frame, $p = (E_n, \vec{p})$ and $p' = (E_n', \vec{p'})$ are the four-momenta of the neutron before and after scattering, $Q^2 = -(k - k')^2$ and $\tilde{Q}^2 = -(p' - p)^2$ are the four-momentum transfers for BDM and neutron, $P_{\mathrm{RFG}}(\vec{p},E_R)$ is the neutron spectral function based on the relativistic Fermi gas (RFG) model and $E_R$ is the removal energy, and $K_{\mathrm{Pauli}}(|\vec{q}|)$ is the Pauli blocking factor depending on the magnitude of the 3-momentum transfer to the neutron $|\vec{q}| = |\vec{p}' - \vec{p}|$. To ensure the scattering is in the QES regime, we only consider scatterings with $|\vec{q}|>350$ MeV~\cite{lin2025detectinglightdarkmatter}. We assume the neutron form factor $F_n(Q^2)$ to be the same as the proton form factor given in Eq.~(\ref{simple_dipole}). The IA scheme assumes the incoming particle resolves a nucleus into a collection of individual nucleons, and the residual nucleus remains unaffected by the knocked-out nucleon during the scattering process~\cite{PhysRevD.72.053005,PhysRevC.77.044311}. Each bound neutron is modeled using the RFG model with identical momentum distributions. The spectral function for the neutrons in the RFG model is~\cite{Bodek:2018lmc}
\begin{align}
    &\notag P_{\mathrm{RFG}}(\vec{p},E_R) \, d^3\vec{p} \, dE_R\\
    &= \frac{\delta(E_R-\bar{E}_R)}{N_F}  p_n ^2 \sin\theta_n  \, dp_n \, d\theta_n \, d\phi_n\, dE_R,
\end{align}
where spherical coordinates $(p_n ,\theta_n,\phi_n)$ are chosen such that the $x-z$ plane corresponds to the plane formed by $\vec{p}\prime $ and $\vec{k} $. The Dirac delta function here constrains $E_R$ to the average value $\bar{E}_R=27.1$ MeV~\cite{Bodek:2018lmc}. The normalization factor is $N_F = \frac{4}{3} \pi p_F^3$, with the Fermi momentum of neutrons in carbon $p_F = 221~\mathrm{MeV}$~\cite{lin2025detectinglightdarkmatter,Bodek:2018lmc}.

After scattering, the knocked-out neutron experiences nuclear effects from the residual nucleus. Pauli blocking is included to suppress the cross section at low three-momentum transfer $|\vec{q}| < 2 p_F$~\cite{Bodek:2018lmc,Tsai:1973py}:
\begin{align}
\label{Eq:Pauli_blocking}
    K_{\mathrm{Pauli}}(|\vec{q}|) 
    = \frac{3}{4} \frac{|\vec{q}|}{p_F} 
      \left[ 1 - \frac{1}{12} \left( \frac{|\vec{q}|}{p_F} \right)^2 \right] ,
\end{align}
where $K_{\mathrm{Pauli}}(q)=1$ if $|q| > 2 p_F$. To account for the final state interaction (FSI) effect, we modify the final KE of the neutron using the nuclear optical potential~\cite{lin2025detectinglightdarkmatter, Bodek:2018lmc, Cai_2020}:
\begin{align}
\label{FSI_energy}
    U_{\mathrm{opt}}\!\left(|\vec{p'}|\right) 
    = \min\!\left[0,\,-29.1 + \frac{40.9}{\mathrm{GeV}^2} \, |\vec{p'}|^2 \right] \ \mathrm{MeV}.
\end{align}
The KE after the neutron escapes the nucleus is then given by
\begin{align}
    \bar{K}'_n = K_n' + U_{\mathrm{opt}}\!\left(|\vec{p'}|\right).
    \label{FSI_i}
\end{align}
We assume that FSI affects only the KE of the knocked-out neutron and not its direction. The corresponding differential cross section becomes
\begin{align}
    \frac{d^2\sigma_{\chi \rightarrow C}}{d\bar{K}'_n \, d\Omega_n'} 
    = \frac{d^2\sigma_{\chi \rightarrow C}}{dK_n' \, d\Omega_n'} 
      \cdot \frac{dK_n'}{d\bar{K}'_n}.
    \label{FSI_ii}
\end{align}

\subsection{\label{Triggered Counts in Detector} Triggered Counts in Detector}

Combining the attenuated BDM flux and the QES differential cross section, we obtain the spectrum of neutrons knocked out in the detector frame by
\begin{align}
\label{differential_counts_eq}
    \frac{dN_{\text{count}}}{d\bar{K}'_n \, d\Omega_n'} 
    = n_C \iiint dt \, d\Omega_\chi \, d\bar{K}_\chi \,
    \frac{d\bar{\Phi}_\chi}{d\bar{K}_\chi \, d\Omega_\chi} \,
    \frac{d^2\sigma_{\chi \rightarrow C}}{d\bar{K}'_n \, d\Omega_n'}(\bar{K}_\chi),
\end{align}  
where $n_C $ is the number of carbon nuclei in the LLSD, $\Omega_\chi$ denotes the incoming BDM direction (with each sky patch treated as an independent source) and $t$ is the time stamp. To reduce computation time, we modify the integrand using Eq.~(\ref{flux_conserve}). Nonetheless, the differential cross section depends on the attenuated BDM KE $\bar{K}_\chi\!\left(K_\chi, z(\Omega_\chi,t)\right)$, where the attenuation depth $z$ varies with both time $t$ and sky patch $\Omega_\chi$ due to the Earth's rotation (see Fig.~\ref{fig:sky_patch}). Hence, we integrate over a 24-hour period for the time integral and multiply by the number of days in the exposure time. For all solid‑angle integrations in this work, we divide the whole sky into 768 equal solid‑angle patches (each of size $7.33^{\circ} \times 7.33^{\circ}$) using the \texttt{healpy} Python package~\cite{Zonca2019,2005ApJ...622..759G}, enabling us to determine the distribution of neutron emission directions in the detector. 

\section{\label{sec_2}Neutron Propagation}
This section discusses the propagation of neutrons through the medium after their emission. As neutrons scatter with protons or nuclei, the produced energetic secondary charged particles — and the electromagnetic showers they induce — can trigger scintillation light, denoted as secondary light in this work. In addition, the residual nucleus, from which the neutron is emitted, may de-excite via $\sim$MeV gamma ray emission or decay to other nuclei products~\cite{Cheng:2024uyj}. The combination of light from residual nucleus de-excitation and secondary light can be reconstructed as the initial vertex, which corresponds to where the neutron is emitted.

After undergoing multiple scatterings and reaching thermal velocities, neutrons are eventually captured by nuclei, emitting gamma rays with characteristic energies. In LS, neutron capture occurs predominantly on hydrogen (99\%) emitting a 2.2~MeV gamma ray and less frequently on carbon (1\%) emitting a 4.9~MeV gamma ray. The mean capture time is approximately 210~$\mu$s~\cite{steven_beacom}. This distinctive gamma-ray emission can be reconstructed as another interaction vertex.

By combining these two reconstructed vertices to form a displacement vector, partial directional information about the neutron emission may be retained. To quantify this effect, we employ Geant4 simulations of the post-emission interactions and account for experimental position‑reconstruction uncertainty by incorporating the vertex resolution.

\subsection{\label{Simulation Setup}Simulation Setup}
Geant4 is a comprehensive Monte Carlo simulation toolkit for modeling particle transport and interactions~\cite{GEANT4:2002zbu,Allison:2006ve,Allison:2016lfl}. It is widely used in nuclear and particle physics experiments to simulate detector responses and interaction processes. In this work, we employ \texttt{geant4-v11.2.2}, using the physics lists \texttt{G4HadronElasticPhysicsHP} and \texttt{G4HadronPhysicsFTFP\_BERT\_HP} to model hadronic interactions involving neutrons. The ``HP" indicates the use of high-precision thermal neutron scattering data, which is particularly suitable for neutrons with KE below 20~MeV~\cite{Allison:2016lfl}. We also include \texttt{G4EmStandardPhysics} to account for EM interactions from secondary particles. The chemical composition of the LS is taken to follow that of the Jiangmen Underground Neutrino Observatory (JUNO) and is summarized in Table~\ref{tab:ls_composition}~\cite{An_2016}. In this work, we do not consider the detector configuration (e.g. geometry) and detector response modeling.
\begin{table}[t!]
\centering
\scriptsize 
\begin{tabular}{|l|c|c|}
\hline
\textbf{Chemicals} & \textbf{Composition} & \textbf{Density (g/cm$^3$)} \\
\hline
Linear alkylbenzene (LAB) & C$_{6}$H$_{5}$C$_{n}$H$_{2n+1}$ & 0.855985 \\
2,5-diphenyloxazole& C$_{15}$H$_{11}$N$_1$O$_1$ & 0.003  \\
p-bis-(o-methylstyryl)-benzene & C$_{24}$H$_{22}$ & $1.5\times 10 ^{-5}$ \\
\hline
\end{tabular}
\begin{tabular}{|l|c|c|}
\hline
\textbf{Elements} & \textbf{Number Density (cm$^{-3}$)} & \textbf{Abundance Ratio} \\
\hline
C & $3.78 \times 10^{22}$ & 37.5\% \\
H & $6.29 \times 10^{22}$ & 62.5\% \\
O & $8 \times 10^{18}$ &  0.008\% \\
N & $ 8 \times 10^{18}$ & 0.008\% \\
\hline
\end{tabular}
\caption{Chemical composition and density of the components in the LLSD, taken to follow JUNO~\cite{An_2016}. Typical values of the alkyl chain length $n$ for LAB range from 10 to 13; $n = 12$ is adopted for the simulation. The number density and atomic abundance ratio of the main elements are also shown.}
\label{tab:ls_composition}
\end{table}

\begin{figure}[!t]
    \centering
    \includegraphics[width=\linewidth]{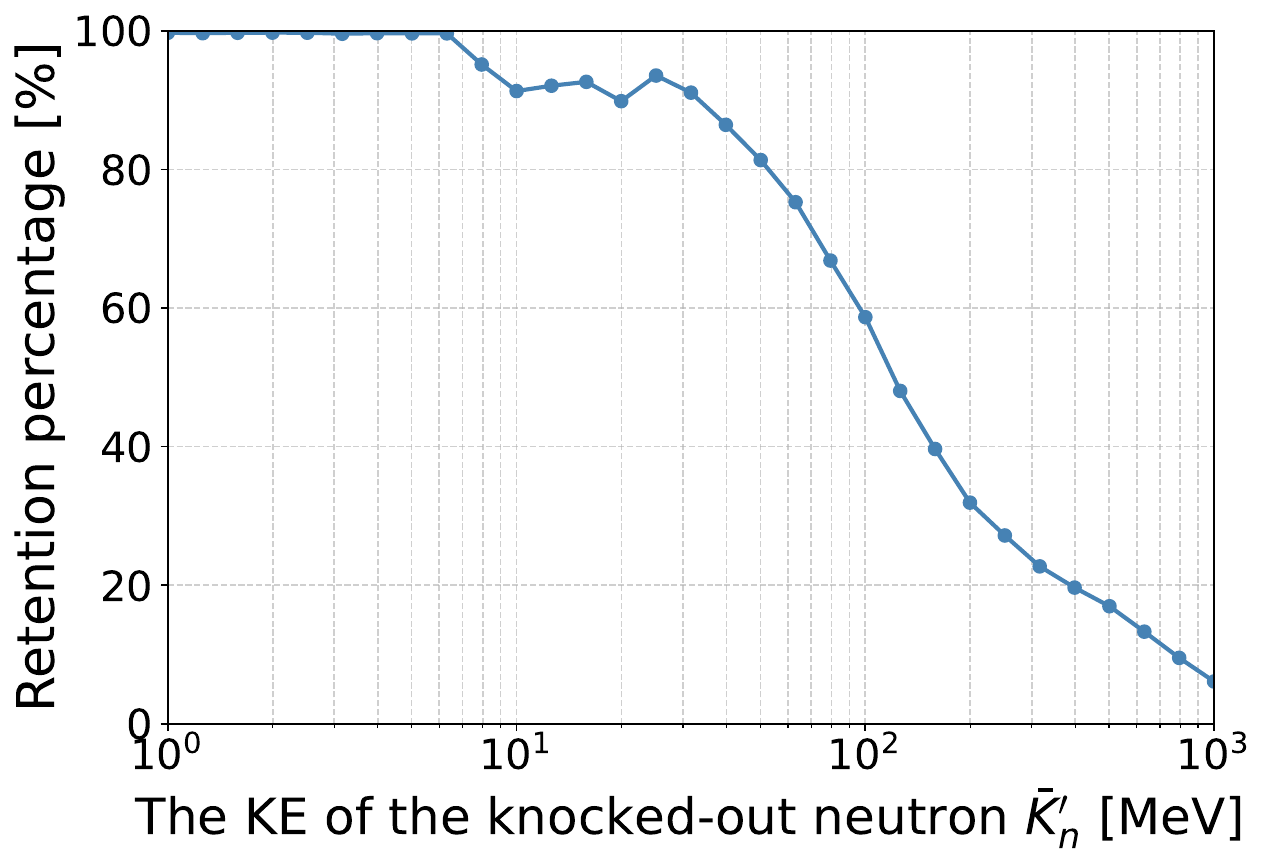}
    \caption{The retention percentage of single-neutron-capture simulation events against the KE of the knocked-out neutron.}
    \label{single_cap}
\end{figure}

Following the configuration of the physics lists, we inject $10^5$ neutrons at the origin $(0,0,0)$ (treated as the position of the residual nucleus) and shoot them along the positive z‑axis (the neutron emission direction) for different KEs. As the KE of the knocked-out neutron $\bar{K}'_n$ increases, inelastic scattering becomes more frequent, generating secondary neutrons that may complicate the primary neutron directional reconstruction. Hence, we record the number of neutrons captured in each event, and select events with only one neutron being captured throughout the simulation. The retention proportions for different knocked-out neutron KE are shown in Fig.~\ref{single_cap}; these are multiplied into the event rates to account for signal loss.
From these single‑neutron‑capture events, we record \\

\noindent i) the deposited energy $E_{\mathrm{secondary}}$ from secondary light;\\
ii) the neutron's capture coordinates $(\mu_x, \mu_y, \mu_z)$.\\

The deposited energy is in the form of scintillation light. Heavier charged particles (such as protons or nuclei) produce less light due to the quenching. The quenched deposited energy is calculated with Birks formula in each simulation step~\cite{Birks:1951boa,Birk_2}
\begin{align}
\label{prompted_deposit_energy}
    \Delta E^{p/N}_{\mathrm{secondary}} = S \left[ 1+k_B \left( \frac{\Delta E}{\Delta d} \right)+C\left(\frac{\Delta E}{\Delta d} \right) ^2\right]^{-1} \Delta E,
\end{align}
where $S$ is the scintillation efficiency which is assumed to be 1, $k_B=0.0098\text{~cm~MeV}^{-1}$ and $C=1.0\times10^{-7}\text{~cm}^2 \text{~MeV}^{-2}$ are the quenching parameters for the chemical composition taken in this work~\cite{vonKrosigk:2013sa}, $\Delta E$ and $\Delta d$ are the energy loss and propagation distance of the charged particle in the simulation step. The quenching effect of electrons and positrons is neglected in this study, i.e. $\Delta E^{e^+/e^-}_{\mathrm{secondary}} = \Delta E$. We sum the deposited energies of all charged leptons, protons and nuclei over all simulation steps to obtain the deposited energy from secondary light for a single event. The de-excitation energy of the residual nucleus depends on the excitation energy left after the knockout and requires a simulation toolkit, e.g. GEMINI++~\cite{charity_2008}, to model the de-excitation products. Including the de-excitation energy shifts the total prompt-energy spectrum to higher energies, but in general changes the prompt spectrum only mildly~\cite{Cheng:2024uyj}. In this work, we include only the deposited energy from secondary light — not the de-excitation energy of the residual nucleus — in the prompt signal, so $E_{\mathrm{prompt}}=E_{\mathrm{secondary}}$.

Events with a single captured neutron are assumed to be tagged as signal candidates with 100\% efficiency. Since the delayed gamma ray from neutron capture is in a time window distinct from the prompt signal, the two vertices can be reconstructed separately. The neutron capture angle $\theta_c$ between the neutron capture position vector and the initial emission direction is calculated via
\begin{align}
\label{n_cap_angle}
\cos\theta_c = \frac{\mu_z}{\sqrt{\mu_x^2 + \mu_y^2 + \mu_z^2}}.
\end{align}
This non-zero angle helps quantify the smearing effect due to neutron diffusion. 
\begin{figure}[htbp]
    \centering
    \includegraphics[width=\linewidth]{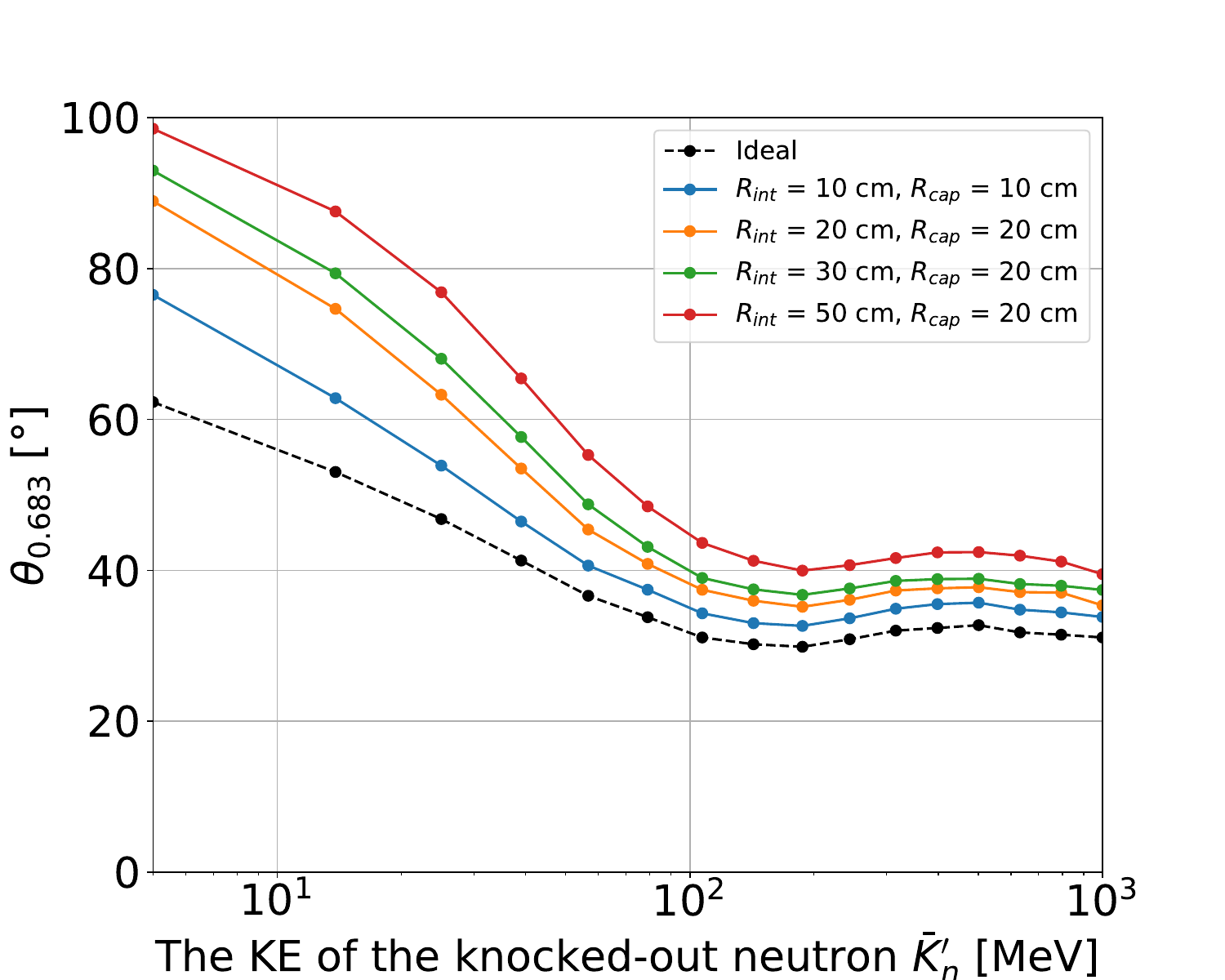}
    \caption{The 68.3\% quantile angle as a function of the KE of the knocked-out neutron, $\bar{K}'_n$, for five different vertex resolution configurations. The ideal case (black dashed) uses the true interaction positions without reconstruction. An optimistic case is shown (blue) for $R_{\text{int}} = 10$~cm and $R_{\text{cap}} =10$~cm. $R_{\text{int}} = 20$~cm and $R_{\text{cap}} = 20$~cm (orange) are adopted for sensitivity estimation in this work. For the other two cases, the vertex resolution for neutron capture points is fixed at $R_{\text{cap}} = 20$~cm, while the vertex resolution for the initial vertex $R_{\text{int}}$ is set to 30~cm (green) and 50~cm (red) for comparison.}
    \label{vertex_resolution}
\end{figure}
\begin{figure*}[htbp]
    \centering
    \includegraphics[width=\linewidth]{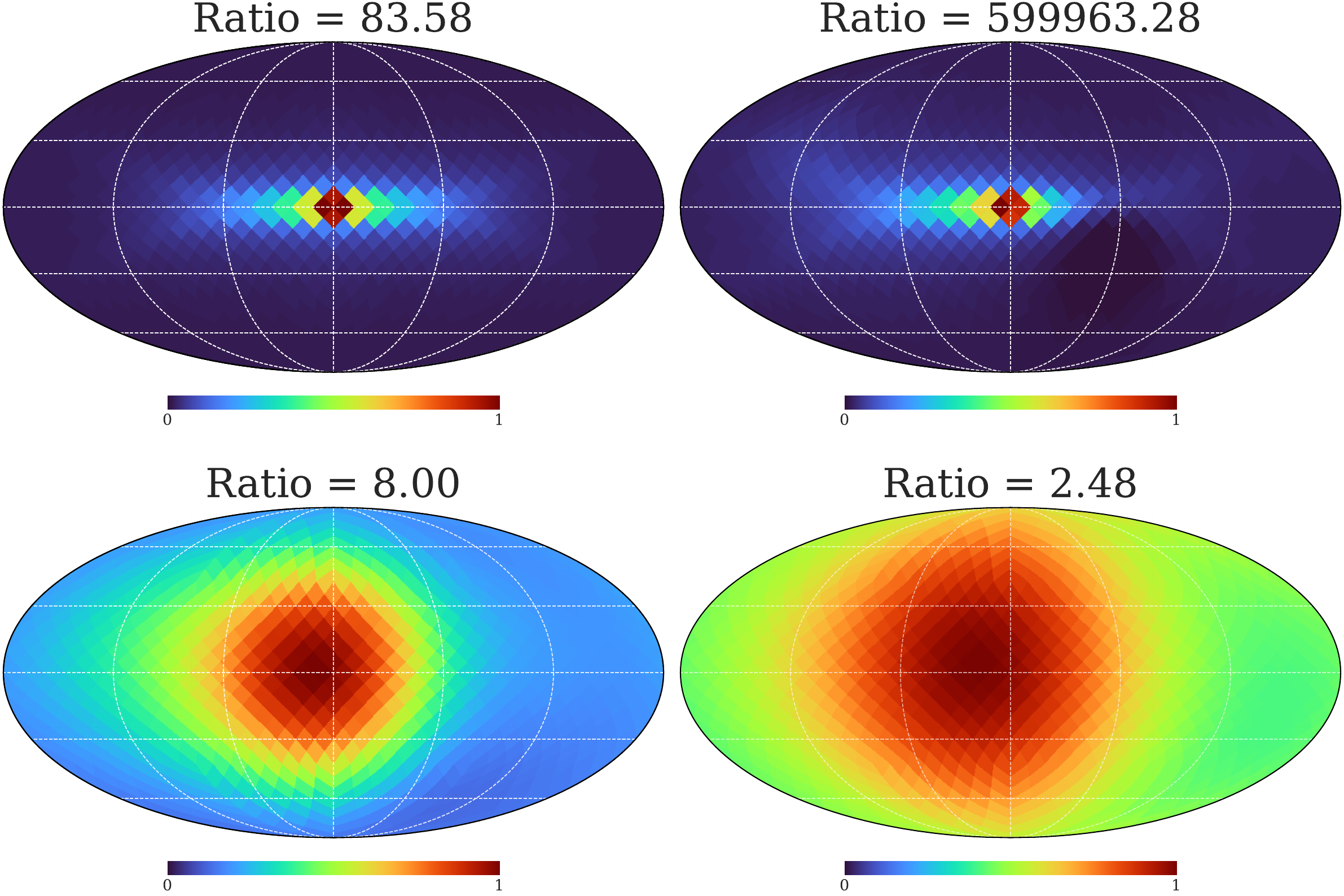}
    \caption{The angular distribution of signal in Galactic coordinates. As an example, we take \( m_\phi = 1~\text{GeV} \), \( m_\chi = 1~\text{MeV} \), and \( \bar{\sigma}_n = 3.71 \times 10^{-34}~\text{cm}^2 \) for this figure. To highlight the contrast difference between the maps, we divide the values of the pixels by the value of the brightest pixel in each panel. The four sky maps correspond to: (top left) the BDM arrival map at Earth before attenuation, (top right) the attenuated BDM arrival map at the detector, (bottom left) the knockout neutron emission map, and (bottom right) the neutron emission map modified by neutron diffusion and vertex reconstruction effect. The ratio between the brightest and dimmest pixels is shown in the title of each sky map.}\label{fig:bdm_neutron_maps}
\end{figure*}

\subsection{\label{Vertex Resolution}Vertex Resolution}
In an LLSD, the vertex of scintillation light is reconstructed via a maximum-likelihood algorithm that fits the recorded photon arrival times and charges from multiple PMTs to pre-calculated probability density functions in order to determine the interaction position~\cite{Liu:2018fpq}. For simplicity, we assume that the reconstructed position is a three‑dimensional random point $\boldsymbol{\lambda
} = (\lambda_x, \lambda_y, \lambda_z)$, normally distributed around the true position $\boldsymbol{\mu} = (\mu_x, \mu_y, \mu_z)$ in Cartesian coordinates, following~\cite{qishanliu2026}. For each simulated interaction vertex $i$ with coordinates $\mu^i \in \{\mu_{x}^i, \mu_{y}^i, \mu_{z}^i\}$, we generate the corresponding reconstructed coordinates $\lambda^i \in \{\lambda_{x}^i, \lambda_{y}^i, \lambda_{z}^i\}$ using the 1D normal probability density function
 \begin{align}
    \label{Eq.prob_distribution}
    f(\lambda^i) = \frac{1}{\sigma \sqrt{2\pi}} \exp\left(-\frac{(\lambda^i - \mu^i)^2}{2\sigma^2}\right),
\end{align}
where $\sigma = R_{\text{vertex}} / \sqrt{\chi^2_{0.683,3}}$ is the standard deviation for each Cartesian coordinate, determined by scaling the 3D vertex resolution $R_{\text{vertex}}$ with the chi-squared value corresponding to a 68.3\% probability with 3 degrees of freedom, $\chi^2_{0.683,3}$~\cite{qishanliu2026}.

For the initial vertex, most of the secondary light emission positions differ from the actual neutron emission point. Including all of the secondary light would bias the reconstruction of the initial vertex, i.e. the neutron emission point. A time cut is therefore suggested once the prompt signal is identified, using only the early secondary light for the reconstruction; see Appendix~\ref{sec:FN_light} for details. The distances of the secondary-light emission points from the neutron emission point are within $\sim$10 cm for $\bar{K}'_n\leq 100$ MeV and within $\sim$20 cm for 100 MeV $<\bar{K}'_n\leq1$ GeV, see Fig.~\ref{fig:distance_and_time}. For the capture vertex, the 2.2 MeV gamma ray from neutron capture is electrically neutral and can travel a distance of $\mathcal{O}(10)$ cm before interacting with the electrons via Compton scattering. Consequently, the scintillation light emission is offset from the true neutron capture point by a displacement comparable to this travel distance. For the vertex reconstruction of scintillation light, the resolution is estimated to be $\sim$10 cm, given that the vertex resolution for positron events at 1 MeV is around 9 cm in JUNO~\cite{Takenaka_2025}. Combining the $\sim10$ cm propagation distance and $\sim 10$ cm vertex reconstruction effect, $R_{\text{int}} = 20$~cm and $R_{\text{cap}} = 20$~cm are adopted in the sensitivity calculation as a conservative choice in this work.

\subsection{\label{smoothing_effect}Simulation Results}

To convert the KE of the knocked-out neutron into prompt energy, we derive a transfer function $F(E_{\mathrm{prompt}}\mid\bar{K}'_n)$ from the simulation results. For each $\bar{K}'_n$, this function gives the normalized probability distribution of the prompt energy $E_{\mathrm{prompt}}$. The expected signal and background event counts, binned as a function of the KE of the knocked-out neutron, $N(\bar{K}'_n)$, are mapped to prompt-energy bins, $N(E_{\mathrm{prompt}})$ via
\begin{align}
\label{Eq:deposit_transfer}
N(E_{\mathrm{prompt}})=\sum_{\bar{K}'_n} N(\bar{K}'_n)\,F(E_{\mathrm{prompt}}\mid\bar{K}'_n).
\end{align}

Using the reconstructed coordinates generated from Eq.~(\ref{Eq.prob_distribution}), we recalculate $\cos\theta_c$ for each event and determine the 68.3\% quantile angle $\theta_{0.683}$ for each $\bar{K}'_n$ bin. This angle represents the range within which 68.3\% of events fall in the distribution of angular differences between the reconstructed direction and the true direction of the neutron~\cite{qishanliu2026}. We use $\theta_{0.683}$ as the standard deviation for the Gaussian smearing effect to smooth the knockout neutron emission map. 

To demonstrate the effect of vertex reconstruction, we consider five cases, as shown in Fig.~\ref{vertex_resolution}, varying the initial vertex resolution $R_{\text{int}}$ and the capture vertex resolution $R_{\text{cap}}$. These five cases are: the ideal case in which both the initial and capture reconstructed positions are the actual positions, an optimistic case with $R_{\text{int}}=10$~cm and $R_{\text{cap}}=10$~cm, and three additional cases with $R_{\text{cap}} = 20$~cm and $R_{\text{int}} = \{20, 30, 50\}$~cm. The smearing effect due to vertex resolution is more severe at low energies because the separation between vertices is comparable to the vertex resolution. In contrast, the increased vertex separation reduces the impact of smearing at higher energies.
\begin{figure*}[t!]
    \centering
    \includegraphics[width=\linewidth]{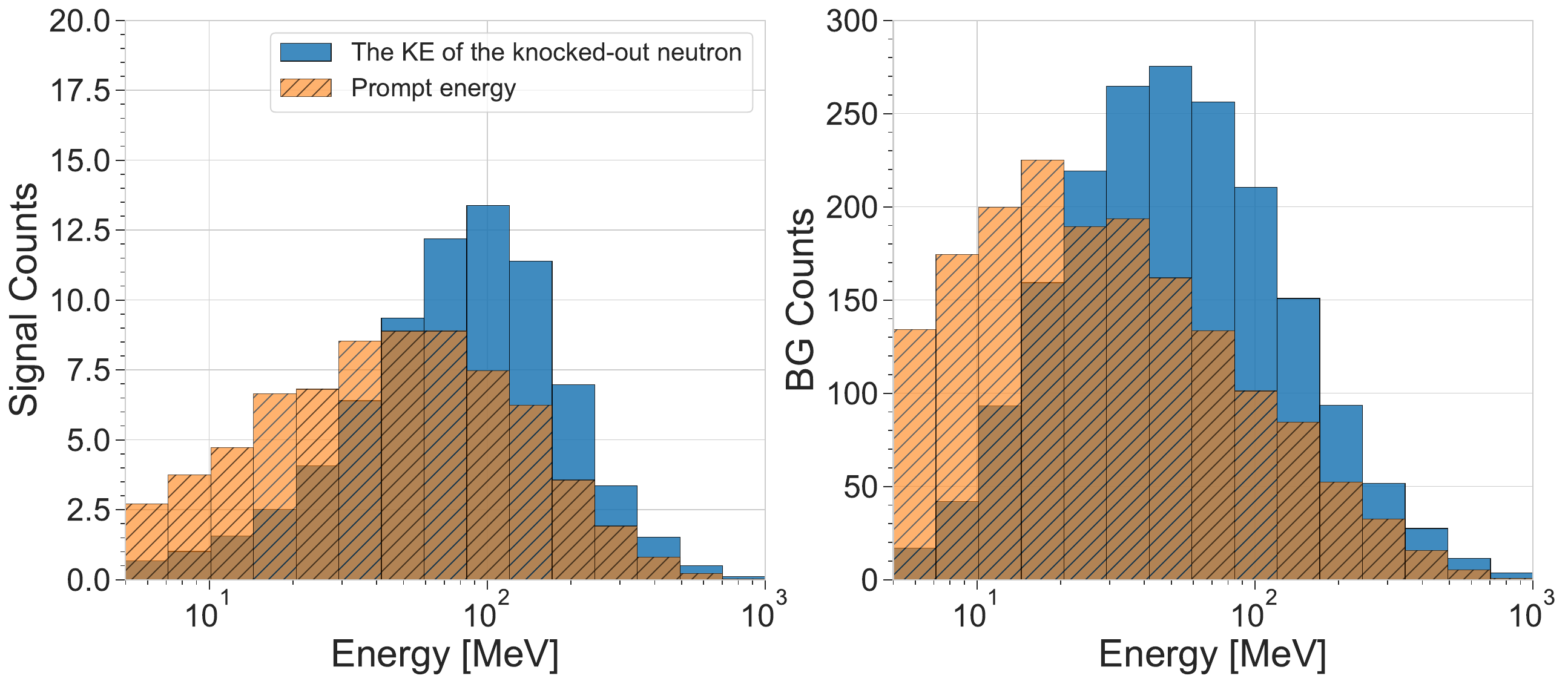}
    \caption{Signal (left) and background (right) counts in LLSD for an exposure of 100~kt$\cdot$yr across different energy bins. Left: We show the KE of the knocked-out neutron (blue) and the prompt energy from Geant4 results (orange). For the signal, we take $m_\phi = 1~\text{GeV}$, $m_\chi = 1~\text{MeV}$, and $\bar{\sigma}_n = 2.56 \times 10^{-35}~\text{cm}^2$ as an example. Right: We show the indistinguishable background from atmospheric neutrino neutral-current quasi-elastic scattering (NCQES) interactions that knock out a neutron, based on the simulation results of GENIE with the RFG model from Ref.~\cite{JieCheng_I,JieCheng_II}.} 
    \label{fig:deposit_signal}
\end{figure*}
Fig.~\ref{fig:bdm_neutron_maps} shows the signal-only BDM map before and after attenuation, together with the signal-only neutron map before and after diffusion and reconstruction for \( m_\phi = 1~\text{GeV} \), \( m_\chi = 1~\text{MeV} \), and \( \bar{\sigma}_n = 3.71 \times 10^{-34}~\text{cm}^2 \). Compared with the unattenuated BDM map (top left), the reconstructed neutron map (bottom right) is much less concentrated toward the GC: the nucleon scattering angle alone reduces the brightest-to-dimmest pixel ratio from 83.58 to 8.00, and diffusion and vertex reconstruction effect reduce it further to 2.48. The extremely high ratio, $\sim6\times 10^5$, for the attenuated BDM map is due to Earth attenuation being too strong in one direction, which makes the denominator of the ratio very small.

\section{\label{sec_3}Sensitivity}
In this section, the sensitivities of the knockout-neutron signals, with and without directional information, are calculated for an LLSD with 100~kt$\cdot$yr exposure. 

\subsection{\label{Signal and Background}Signal and Background} 
Eq.~(\ref{differential_counts_eq}) is integrated to obtain the number of events in each $\bar{K}'_n$ bin for each pixel bin. Combining this with the single-neutron capture retention proportion (see Fig.~\ref{single_cap}), transfer function (Eq.~(\ref{Eq:deposit_transfer})) and the angular smoothing effect described in Sec.~\ref{smoothing_effect}, we obtain the expected number of events in each prompt-energy bin for each pixel bin. The upper prompt-energy bound is set to \(1\ \mathrm{GeV}\) because the retention percentage drops sharply when the KE of the knocked-out neutron exceeds \(1\ \mathrm{GeV}\), see Fig.~\ref{single_cap}. The lower bound is set to \(5\ \mathrm{MeV}\) to ensure sufficient light for vertex reconstruction. For each of the 768 equal-solid-angle pixel bins, we use 15 equal log-scale prompt-energy bins as a conservative choice given the excellent energy resolution an LLSD can reach.

The prompt-delayed coincidence requirement strongly suppresses backgrounds consisting of single, isolated energy depositions. The relevant background events share the same signature as the targeted signals, which consist of a prompt signal together with a time‑delayed gamma ray from neutron capture. Spallation neutrons are free neutrons knocked out of atomic nuclei by high-energy cosmic muons. These neutron capture events may pair up with accidental prompt events, such as de-excitation from radioactive cosmogenic nuclei inside the LS, and form a background event. For example, the muon rate in JUNO is about 5~Hz~\cite{nairat2025newspallationbackgroundrejection}, and the rate of neutron-capture events from muon spallation is about 1.6 Hz~\cite{Takenaka_2025}. The water buffer provides a muon veto that reduces the background due to muons entering the LS and knocking out neutrons. However, spallation neutrons produced in the surrounding rock can still penetrate the buffer and reach the scintillator, possibly creating accidental prompt-delayed pairs. To mitigate this background, fiducial‑volume cuts can be applied in the LLSD so that neutron production is effectively confined to the scintillator region under the given setup~\cite{JUNO:2022lpc,steven_beacom}. 

Another possible background arises from inverse beta decay (IBD) induced by atmospheric and reactor electron antineutrinos \((\bar{\nu}_e)\) interacting with free protons in the LS. IBD events consist of a prompt signal from an energetic positron and a delayed neutron-capture signal, which appears to be very similar to BDM-induced prompt-delayed events in LLSD. For IBD, a neutron with \(\bar{K}'_n = 5~\mathrm{MeV}\) would be accompanied by a positron with KE of order \(\sim 50~\mathrm{MeV}\). The prompt energy in IBD events is therefore dominated by the positron, whereas in our signal it is dominated by secondary light that is quenched. Using pulse-shape discrimination (PSD) between positron and secondary-charged-nuclei-induced waveforms, the prompt signal can be distinguished~\cite{Cheng:2023zds}. Charged-current quasi-elastic (CCQE) interactions of atmospheric \(\bar{\nu}\) on bound protons in \({}^{12}\mathrm{C}\) can give the same coincidence signature and are rejected by the same PSD argument. Hence, IBD and CCQE events are not considered as background in this work.

In this work, for neutrino energies $E_\nu$ between 100 MeV and 10 GeV, we consider only the indistinguishable background from the dominant atmospheric neutrino NCQES on carbon:
\begin{align}
\label{neutrino_reaction_EQ}
    \nu + {}^{12}\mathrm{C} \rightarrow \nu + n + {}^{11}\mathrm{C}^*.
\end{align}
Since both neutrino and BDM are invisible inside the detector, the outgoing products of Eqs.~(\ref{reaction_EQ}) and (\ref{neutrino_reaction_EQ}) are the same. This work is restricted to the one-particle-one-hole case to remain consistent with the single-neutron-capture requirement. 

Neutrons induced by atmospheric-neutrino NCQES are assumed to be isotropic in Galactic coordinates, due to the Earth’s rotation and the mostly isotropic production of atmospheric neutrinos. The energy-dependent background distribution is modeled using simulation results from~\cite{JieCheng_II}. Similar to the signal treatment, we model their neutron spectrum by accounting for the single-neutron-capture retention proportion in Fig.~\ref{single_cap}, the FSI effect in Eq.~(\ref{FSI_i}) and Eq.~(\ref{FSI_ii}), and the transfer function in Eq.~(\ref{Eq:deposit_transfer}). 

\begin{figure*}[!t]
    \centering
    \includegraphics[width=\textwidth]{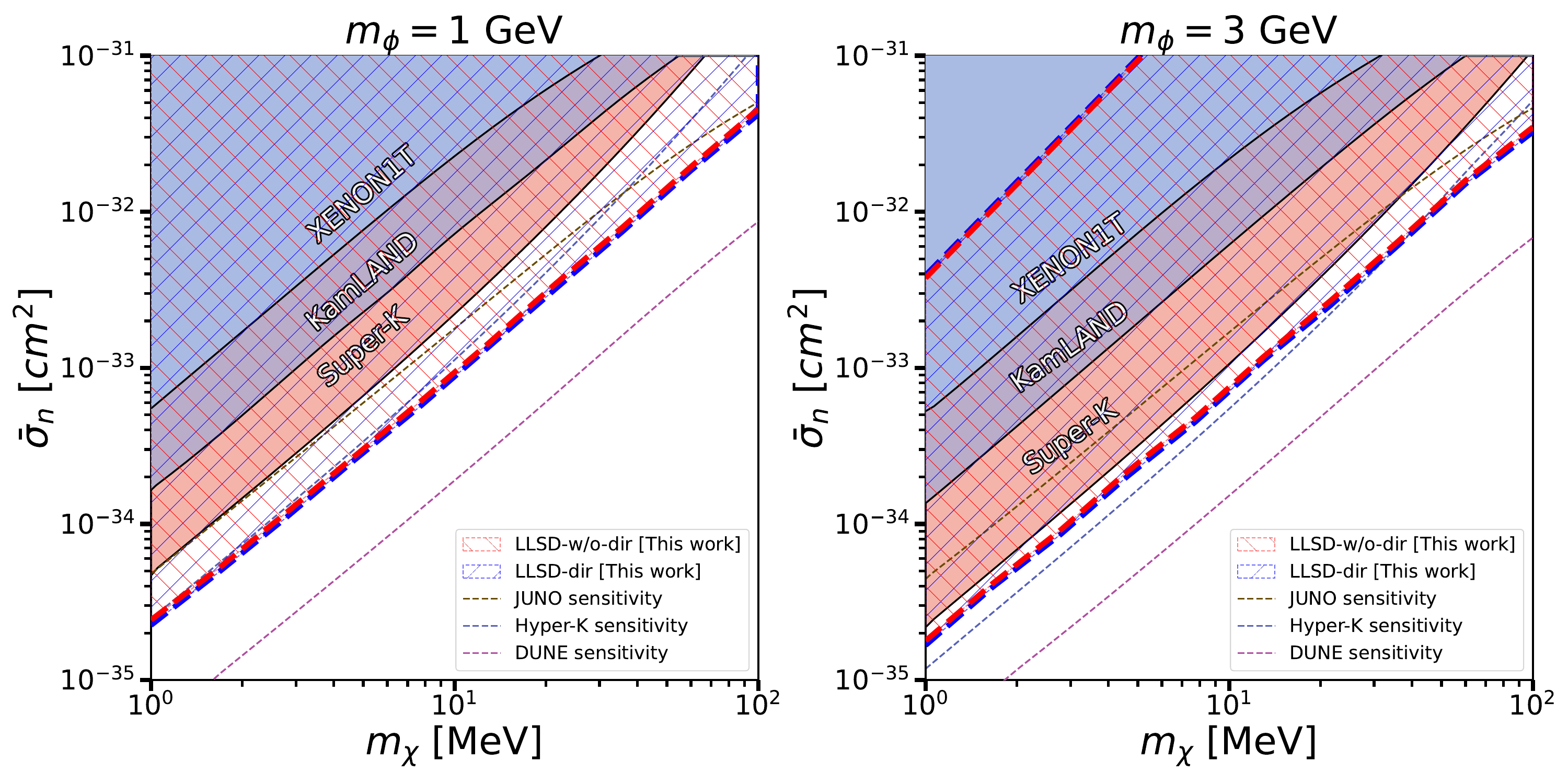}
    \caption{The estimated 95\% CL sensitivity limits on DM-nucleon scattering cross section for a 100 kt$\cdot$yr LLSD of this work compared with the previous results including constraints (solid regions) of XENON1T, KamLAND, Super-K, and sensitivities (dashed lines) of JUNO, Hyper-K, DUNE from Ref.~\cite{Ema_2021}. The left panel corresponds to mediator mass $m_\phi=1\,\text{GeV}$ and the right panel to $m_\phi=3\,\text{GeV}$. The LLSD results of this work are displayed as hashed bands: sensitivities with (without) directional information in blue (red). The upper ceilings of the previous results~\cite{Ema_2021} are omitted to avoid overcrowding the plot.}
    \label{fig:Constraint_plot}
\end{figure*}

To illustrate the Earth attenuation effect, the knockout neutron spectra for both signal (with different $\bar{\sigma}_n$ for $m_\phi = 3$~GeV and $m_\chi = 1$~MeV) and background are shown in Fig.~\ref{fig:neutron_spectrum}. The signal-plus-background sky maps for different $\bar{\sigma}_n$ at $m_\phi = 3$~GeV and $m_\chi = 1$~MeV are shown in Fig.~\ref{fig:neutron_sky_map}. An interesting observation is that the Earth attenuation effect can shift the brightest spot of the signal away from the GC, depending on the geographic location of the detector. The estimated signal for $m_\phi = 1~\text{GeV}$, $m_\chi = 1~\text{MeV}$, and $\bar{\sigma}_n = 2.56 \times 10^{-35}~\text{cm}^2$, together with the atmospheric neutrino NCQES background used in this work, is shown in Fig.~\ref{fig:deposit_signal}, with $\bar{K}'_n$ binning and prompt-energy binning.

\subsection{\label{Estimated Expected Constraint} Expected Sensitivity}
The joint likelihood across $N$ bins is
\begin{align}
    L = \prod_{i=1}^N P_i(k_i \mid \lambda_i),
\end{align}
where $P_i(k_i \mid \lambda_i)$ is the probability of observing $k_i$ events given the expected number $\lambda_i$. This probability is modeled with a Poisson distribution,
\begin{align}
    P_i(k_i \mid \lambda_i) = \frac{\lambda_i^{k_i} e^{-\lambda_i}}{k_i!}.
\end{align}
To obtain the one‑sided 95\% confidence‑level (CL) sensitivity limits on the spin‑independent DM–nucleon scattering cross section $\bar{\sigma}_n$, the log-likelihood ratio is employed as the test statistic~\cite{Cowan:2010js}
\begin{align}
    TS = -2 \ln \frac{L(N_{\text{BDM}})}{L(N_{\text{BDM}}=0)},
\end{align}
where $L(N_{\text{BDM}})$ is the likelihood evaluated for a given BDM signal strength, using the observed counts $k_i = N_{\text{BG}}^i$ (fixed to the background-only expectation) and the expected counts (sum of background and BDM signal) $\lambda_i = N_{\text{BG}}^i + N_{\text{BDM}}^i$ in each bin. In contrast, $L(N_{\text{BDM}}=0)$ is the background-only likelihood, where both the expected and observed counts are purely from background, i.e. $\lambda_i = k_i = N_{\text{BG}}^i$. Setting $TS = 2.71$ yields the one‑sided 95\% CL sensitivity for the detector. 

In Fig.~\ref{fig:Constraint_plot}, we present the 95\% CL sensitivity limits, with and without directional information, for an LLSD with an exposure of $100$~kt$\cdot$yr. The limits are derived from the reconstructed knockout neutron signals produced by BDM for mediator masses $m_\phi = 1$ and $3$~GeV. 

Without directional information, all 768 pixel bins are combined into a single bin without directionality (w/o-dir), resulting in a total of $1\,\text{(w/o-dir bin)} \times 15\,\text{(energy bins)}$. When directional information is included, the analysis uses the full set of $768\,\text{(equal‑solid‑angle pixel bins)} \times 15\,\text{(energy bins)}$. The ceiling of the sensitivity curve at $m_\phi=3$ GeV appears because the Earth attenuation becomes so severe at high cross section — where the number of scatterings grows rapidly — that the incoming BDM lacks sufficient KE to knock out bound neutrons. In addition, for the same $\bar{\sigma}_n$, the Earth attenuation becomes more pronounced with a larger mediator mass.

The key result of this study is that, although the knocked-out neutron retains partial directional information about the incoming BDM (see Fig.~\ref{fig:bdm_neutron_maps}), incorporating this directional information improves the sensitivity only mildly. Finally, we conclude that the only way to enhance directional sensitivity beyond the without-directionality case is to make the signal events more sharply concentrated toward the GC. However, the inherent directionality of the CRBDM is limited by the DM distribution itself, and the neutron emission directions are further smeared by the scattering angle of the knockout; solely using neutron directionality therefore cannot improve the constraint by much.

The limits are also compared with those of Ref.~\cite{Ema_2021}, which considered CRBDM elastic scattering off free protons in hydrogen rather than off bound neutrons. The neutron channel, with its prompt–delayed coincidence search, has less background than the free-proton channel with only prompt signals. Also, the number of bound neutrons is about 3.6 times the number of free protons, see Table~\ref{tab:ls_composition}. In addition, the signal acceptance for the neutron channel covers a wide range of the KE of the knocked-out neutron, see Fig.~\ref{single_cap}. Hence by exploiting the neutron channel, the LLSD can achieve a sensitivity that surpasses JUNO's sensitivity obtained with the free-proton channel, assuming the same exposure~\cite{Ema_2021}. Moreover, the LLSD at $m_\phi = 1~\mathrm{GeV}$ is comparable with the level expected for the next-generation detector Hyper-Kamiokande (Hyper-K) with 5 years of data, which is scheduled to begin operation around the end of 2027~\cite{Hyper-Kamiokande:2025asb}. In this framework and under the assumptions of Ref.~\cite{Ema_2021}, the Deep Underground Neutrino Experiment (DUNE) still provides the strongest projected sensitivity for 5 years of data.

\section{\label{sec_4} Discussion}
Even though the CRBDM source is focused toward the GC direction, the reconstructed neutron signals are smoothed out by the nucleon scattering angle. We have tested the case with diffusion removed and the initial neutron emission direction perfectly reconstructed, yet the improvement remains insignificant since the scattering-angle smearing is still present. It is concluded that, since the initial nucleon scattering angle already smears the map substantially, solely using nucleon directionality cannot provide a significant improvement.

Two optimistic assumptions are made in this work that are likely to overestimate the effect from directionality compared to reality: (1) the BDM direction remains unchanged after each scattering during Earth attenuation, and (2) the knockout direction of the neutron is not affected by FSI. These two assumptions overestimate the event-distribution contrast of the sky map, whereas the actual contrast should be even lower than in this work's simulation results.

A simple test has also been conducted by assuming that the BDM directionality can be perfectly reconstructed, to check whether the inherent directionality of the CRBDM itself is sufficient; see Appendix~\ref{sec:max_improvement} for details. For BDM that originates from DM annihilation in the GC~\cite{Agashe:2014yua,BetancourtKamenetskaia:2025noa}, the J-factor with a $\rho_\chi^2$ dependence of the l.o.s. integral can provide a modest improvement on the angular contrast over the CRBDM case. In addition, for a fixed annihilation cross section, the directionality improvement for $\bar{\sigma}_n$ is squared relative to this work's result, see Appendix~\ref{sec:GC-BDM} for details. The same argument for GC-BDM can be applied to DM--annihilation processes that emit neutrinos~\cite{Kumar:2015nja,IceCube:2015rnn,Arguelles:2019ouk,Super-Kamiokande:2020sgt,Bell:2020rkw,Buckley:2022btu,JUNO:2023vyz,Granelli:2026bem}. Hence, exploring the annihilation channel may give a modestly better improvement from directionality.

\section{\label{Conclusion} Conclusion}
While halo DM may be too light and lack sufficient KE to generate detectable signals in current detectors, DM can be boosted through certain mechanisms to relativistic energies. LLSDs offer low energy thresholds and kiloton-scale target masses, making them suitable for probing such events. In this work, we model the CRBDM flux using a leaky-box description and the LIS CR flux, and include attenuation effects before BDM reaches an underground LLSD. The neutron knockout rate is calculated in the QES regime, and Geant4 simulations are employed to study neutron propagation, showing that reconstructed capture and emission vertices can preserve partial information about the BDM arrival direction.

We consider the indistinguishable atmospheric-neutrino NCQES background and estimate the 95\% CL limits for a 100 kt$\cdot$yr LLSD, both with and without directional information. The results indicate that, for the same exposure LLSD, the neutron-knockout channel yields stronger constraints than the free-proton channel considered in Ref.~\cite{Ema_2021}. At $m_\phi = 1~\mathrm{GeV}$, the sensitivity reaches the level expected for Hyper-K with 5 years of data, while DUNE still provides the strongest projected sensitivity in this framework~\cite{Ema_2021}. However, directional information from the CRBDM mechanism does not yield a significant improvement (e.g. $\sim3\%$ on $\bar{\sigma}_n$ for $m_\chi = 1$~MeV, $m_\phi = 3$~GeV), as the inherent directionality of the CRBDM is smoothed out by the nucleon scattering angle. Although the improvement from directionality is modest, it comes at essentially no cost, as it relies only on the low-energy vertex reconstruction already available in LLSDs. We therefore encourage LLSD experiments to exploit directionality in DM searches and continue to explore avenues for better event reconstruction techniques for directionality. A larger directional gain is expected for channels whose flux is more sharply concentrated toward the GC — such as DM annihilation into BDM or neutrinos, whose J-factor ($\propto \rho_\chi^2$) provides stronger inherent directionality. 

\section{\label{Acknowledgment} Acknowledgments}
SSHT would like to acknowledge helpful conversations with Chingam Fong, Kelly K. K. Chan, and Cheuk Nam Chong. KCYN and SSHT are supported by Croucher Foundation, RGC grants (24302721, 14305822, 14308023), Joint NSFC/GRC grant (N CUHK456/22), and NSFC grant 12322517.
KC is supported by the fund from the Institute for Basic Science (IBS-R016-Y2, IBS-R016-R1). This work is supported in part by the National Natural Science Foundation of China under grant No. 12342502.

\appendix
\renewcommand{\thefigure}{A.\arabic{figure}}
\renewcommand{\theHfigure}{A.\arabic{figure}} 
\setcounter{figure}{0}

\begin{figure}[!htbp]
    \centering
    \includegraphics[width=\linewidth]{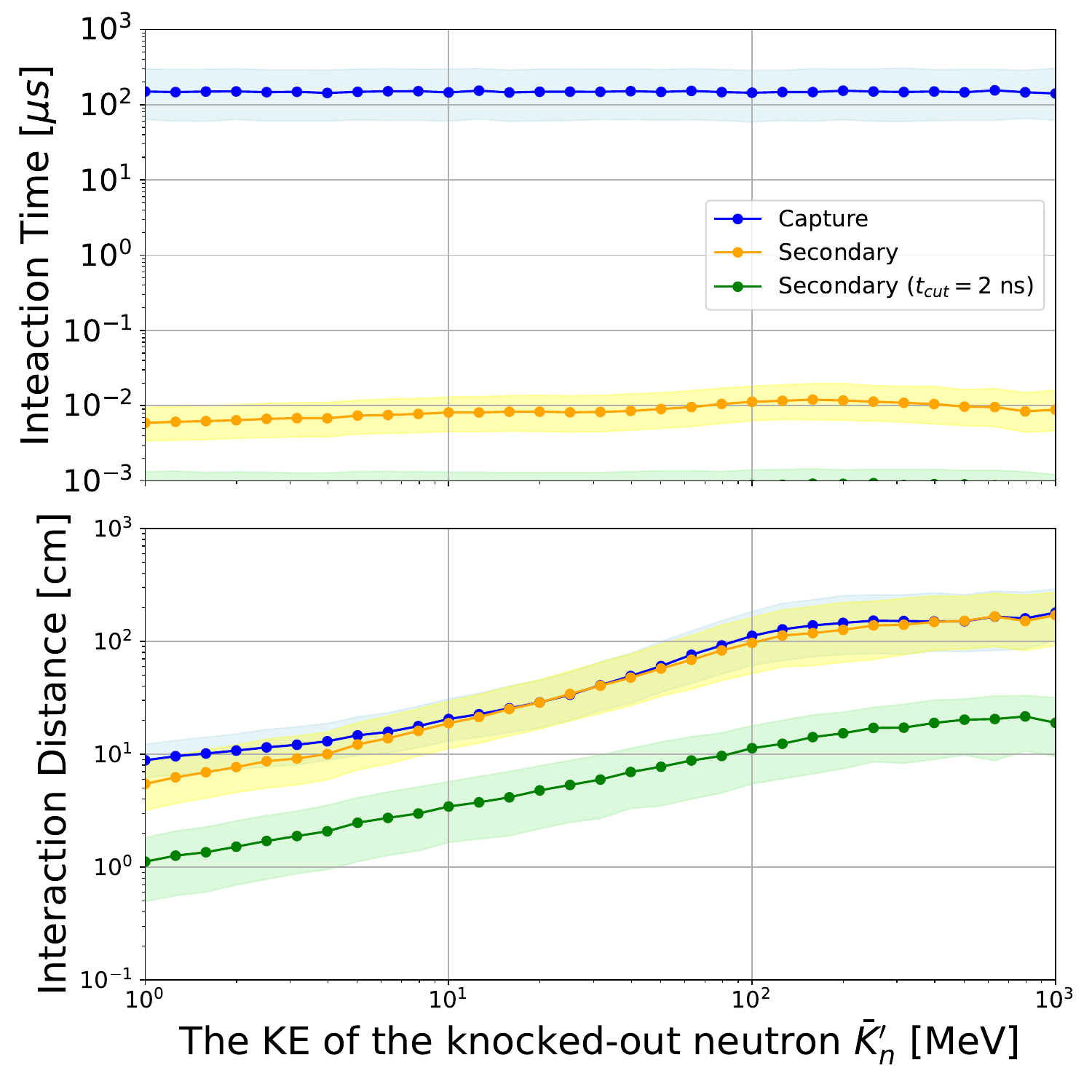}
    \caption{The figure displays the median (solid line) and the 25\%–75\% quantile range (colored region) for both the interaction time and the distance, measured since the moment when the neutron is shot in Geant4 simulation. We record secondary charged particles that undergo ``ionization'' and treat them as sources of scintillation light, labeled as ``Secondary''. The three curves correspond to where the neutron is captured (blue), scintillation light triggered by the secondary charged particles before adding a time cut (yellow) or with a 2 ns time cut (green).}
    \label{fig:distance_and_time}
\end{figure}

\begin{figure}
    \centering
    \includegraphics[width=\linewidth]{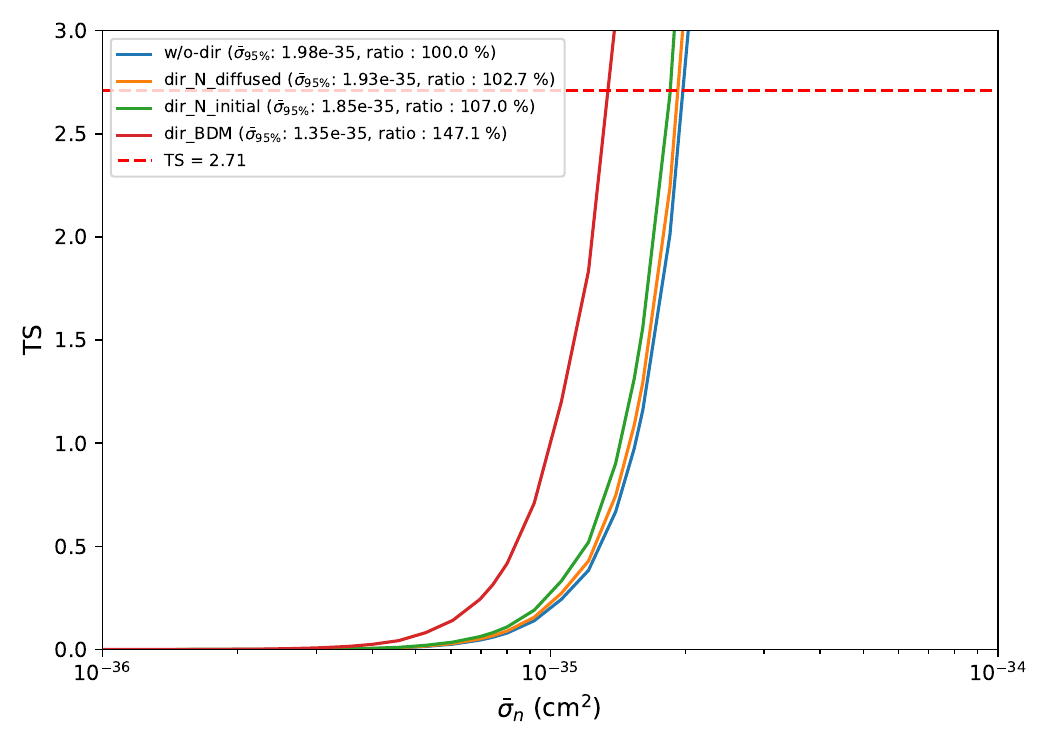}
    \caption{Test statistic versus $\bar{\sigma}_n$ for $m_\chi = 1~\mathrm{MeV}$, $m_\phi = 3~\mathrm{GeV}$, and pixel size $7.33\degree \times 7.33\degree$, for four cases: without directionality (blue); neutron directionality with diffusion and vertex reconstruction effect (orange); neutron directionality with the initial-knockout direction (green); and the BDM directionality (red). The values $\bar{\sigma}_{95\%}$ in the legend are the $\bar{\sigma}_n$ at which $\mathrm{TS} = 2.71$, the one-sided 95\% CL sensitivity (red dashed horizontal line). The ratios shown in the legend are taken relative to the without-directionality case and quantify the sensitivity gain from directionality.}
    \label{fig:TS_against_cross}
\end{figure}

\section{\label{sec:FN_light} How secondary light affects the initial vertex reconstruction}
The KE of the knockout neutron is transferred to secondary particles very quickly, long before the neutron is captured. The emission points for secondary light, even though triggered within tens of nanoseconds, can extend to distances far away from the residual nucleus. If all of these light sources are included for vertex reconstruction, the initial vertex may be significantly displaced from the actual neutron emission point. In principle, this issue can be mitigated by applying a time cut. In Fig.~\ref{fig:distance_and_time}, we apply a time cut of $t_{\text{cut}} = 2$~ns, based on the timing resolution of the PMTs in a JUNO-like LLSD~\cite{An_2016}, once the neutron is emitted. After this time cut, the distance reached by secondary particles is reduced by approximately an order of magnitude compared to the neutron capture position. By selecting the early light emission from the prompt signal, the vertex resolution should be significantly improved.

\section{\label{sec:max_improvement} Why the directional improvement is mild}
To check whether the BDM sky map itself contains enough contrast for directionality to help, we examine a non-physical scenario in which the direction of each incoming BDM particle is assumed to be perfectly reconstructed, so that the event map follows the D-factor distribution Eq.~(\ref{D_factor}). This is the strongest inherent directionality the CRBDM mechanism can provide and therefore sets the best improvement. Figure~\ref{fig:TS_against_cross} compares four cases of increasing realism. Even in this idealized limit, the 95\% CL limit on $\bar{\sigma}_n$ improves by only $\sim$50\% relative to the without-directionality case; restoring the initial neutron-knockout angle reduces the improvement to $\sim$7\%, and further including diffusion and vertex reconstruction effect bring it to only $\sim$3\%. The directionality is thus mostly lost once the nucleon scattering angle is included — not by diffusion — confirming that the failure of neutron directionality is intrinsic and shared with any knocked-out nucleon, proton included. 

\section{\label{sec:GC-BDM} Comparison with the GC-BDM annihilation channel}
Other than CRBDM, BDM flux may also originate from DM annihilation in the GC~\cite{Agashe:2014yua,BetancourtKamenetskaia:2025noa}. This BDM flux angular distribution then follows a J-factor, i.e.\ a line-of-sight integral proportional to $\rho_\chi^2$, which concentrates the flux more sharply toward the GC than the D-factor of the CRBDM case. The larger contrast can potentially increase the gain from directional analysis.

Since the CRBDM event rate scales as $\bar{\sigma}_n^2$, any gain from directionality translates into only a square-root improvement in the $\bar{\sigma}_n$ limit. The annihilation rate instead scales as $\bar{\sigma}_n\,\braket{\sigma v}$, so for fixed $\braket{\sigma v}$ the directional gain is transferred linearly to $\bar{\sigma}_n$. For example, with a J-factor sky map for $m_\chi = 1~\mathrm{MeV}$, $m_\phi = 3~\mathrm{GeV}$ and fixed $\braket{\sigma v}$, the 95\% CL limit on $\bar{\sigma}_n$ is estimated to improve by $\sim20\%$ over the without-directionality case---modestly better than for CRBDM because of the more concentrated distribution.
\begin{figure}[h]
\includegraphics[width=\linewidth]{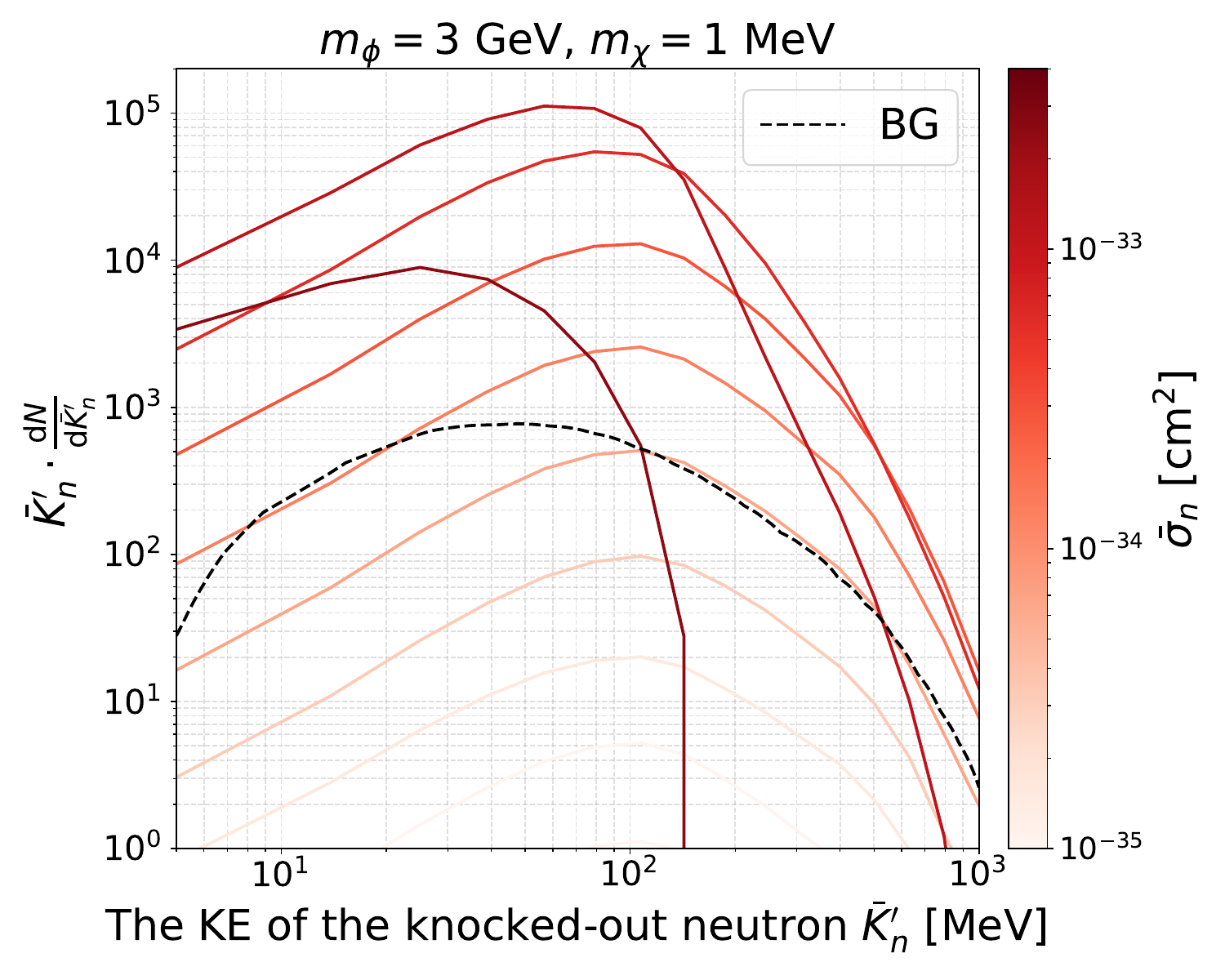}
    \caption{The knockout neutron spectra (red solid) and background (black dashed) for a $100$~kt$\cdot$yr LLSD for different values of the cross section $ \bar{\sigma}_n $, with $ m_\phi = 3\,\text{GeV} $ and $ m_\chi = 1\,\text{MeV} $.}
    \label{fig:neutron_spectrum}
\end{figure}

\begin{figure*}[tbp]
    \centering
    \includegraphics[width=\linewidth]{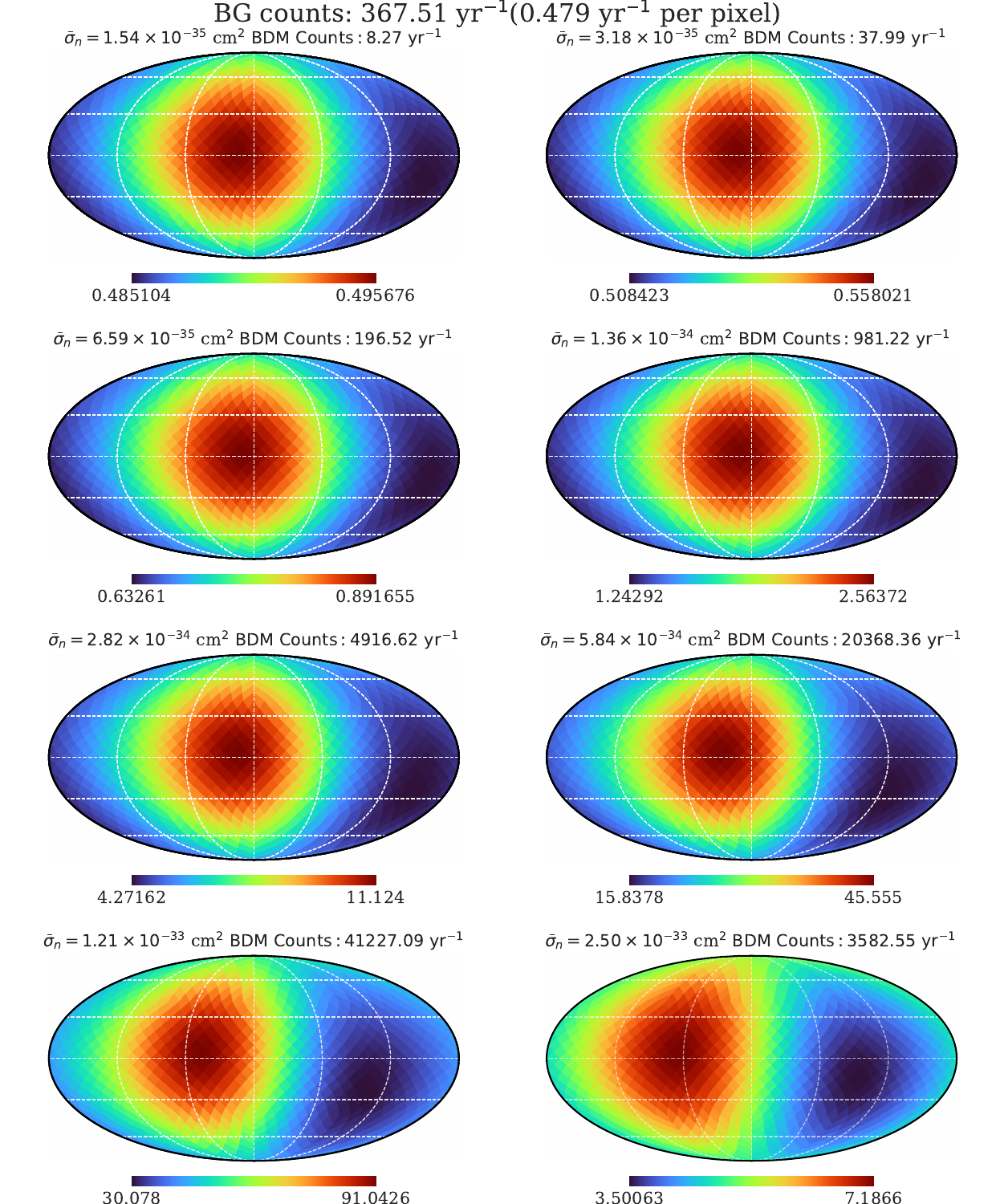}
    \caption{The neutron map (signal + background) after neutron diffusion for different values of the cross section $ \bar{\sigma}_n $, with $ m_\phi = 3\,\text{GeV} $ and $ m_\chi = 1\,\text{MeV} $. The LLSD is taken to be located in the Northern Hemisphere at a latitude of approximately $22^\circ$\,N and about $700~\mathrm{m}$ underground.}
    \label{fig:neutron_sky_map}
\end{figure*}

\clearpage
\bibliography{bib.bib}

\end{document}